\documentclass[3p,sort&compress]{elsarticle}

\usepackage{amssymb}

\usepackage[dvipdfmx,hidelinks,colorlinks=true,linkcolor=black, citecolor=blue, urlcolor=blue]{hyperref}
\usepackage{amsmath}
\usepackage{times}
\usepackage{braket}
\usepackage{multirow}
\usepackage{here}
\usepackage{tabularx}
\newcolumntype{C}{>{\centering\arraybackslash}X}
\usepackage{ulem}
\usepackage{color}

\newcommand{\mrm}{\mathrm}

\definecolor{verde}{rgb}{0.0,0.7,0.0}

\journal{Nuclear Physics A}

\begin{document}

\begin{frontmatter}



\title{Investigation of spatial manifestation of $\alpha$ clusters in $^{16}\mrm{O}$\\ via $\alpha$-transfer reactions}


\author[Fukui]{Tokuro Fukui\corref{cor}}
\ead{fukui@na.infn.it}
\address[Fukui]{Istituto Nazionale di Fisica Nucleare, Sezione di Napoli, Napoli 80126, Italy}
\cortext[cor]{Corresponding author.}

\author[Enyo]{Yoshiko Kanada-En'yo}
\address[Enyo]{Department of Physics, Kyoto University, Kyoto 606-8502, Japan}

\author[Ogata1,Ogata2]{Kazuyuki Ogata}
\address[Ogata1]{Research Center for Nuclear Physics, Osaka University, Ibaraki, Osaka 567-0047, Japan}
\address[Ogata2]{Department of Physics, Osaka City University, Osaka 558-8585, Japan}

\author[Suhara]{Tadahiro Suhara}
\address[Suhara]{Matsue College of Technology, Matsue, Shimane 690-8518, Japan}

\author[Taniguchi]{Yasutaka Taniguchi}
\address[Taniguchi]{Department of Information Engineering, National Institute of Technology, Kagawa College, Mitoyo, Kagawa 769-1192, Japan}

\begin{abstract}
 Recently, we have determined surface distributions of $\alpha$ clusters in the ground state of $^{20}\mrm{Ne}$
 from $\alpha$-transfer cross sections, without investigating the properties of its excited states.
 In this paper we extend our comprehension of $\alpha$-cluster structures in excited states of nuclei through reaction studies.
 In particular we focus on $^{16}\mrm{O}$, for which attention has been paid to advances of structure theory
 and assignment regarding $4^+$-resonance states.
 We study the surface manifestation of the $\alpha$-cluster states in both the ground and excited states of $^{16}\mrm{O}$
 from the analysis of the $\alpha$-transfer reaction $^{12}\mrm{C}(^6\mrm{Li},d)^{16}\mrm{O}$.
 The $\alpha$-transfer reaction is described by the distorted-wave Born approximation.
 We test two microscopic wave functions as an input of reaction calculations.
 Then a phenomenological potential model is introduced to clarify the correspondence
 between cluster-wave functions and transfer-cross sections.
 Surface peaks of the $\alpha$-wave function of $^{16}\mrm{O}(0^+)$ are sensitively probed by transfer-cross sections at forward angles,
 while it remains unclear how we trace the surface behavior of $^{16}\mrm{O}(4^+)$ from the cross sections.
 From inspection of the cross sections at forward angles, we are able to specify that
 the $\alpha$-cluster structure in the $0_1^+$ and $0_2^+$ states
 prominently manifests itself at the radii $\sim 4$ and $\sim 4.5$~fm, respectively.
 It is remarkable that the $4_1^+$ state has the $^{12}\mrm{C}+\alpha$-cluster component
 with the surface peak at the radius $\sim 4$ or outer, whereas the $^{12}\mrm{C}+\alpha$-cluster component
 in the $4_2^+$ state is found not to be dominant.
 The $4_2^+$ state is difficult to be interpreted by a simple potential model assuming the $^{12}\mrm{C}+\alpha$ configuration only.
\end{abstract}

\begin{keyword}
 Transfer reaction \sep Cluster structure



\end{keyword}

\end{frontmatter}


\section{Introduction}
\label{intro}
The formation of cluster structures, together with that of mean fields, is a fundamental aspect in nuclear many-body dynamics.
Theoretically, various cluster states are predicted in excited states of light-stable nuclei as well as in
$sd$-shell or unstable nuclei (see, for instance,
Refs.~\cite{PhysRev.52.1083,PhysRev.52.1107,PhysRev.96.378,BRINK1970143,Suzuki:1976zz,Suzuki:1976zz2,fujiwara80} and references therein).
Let us consider as an example the double magic nucleus $^{16}\mrm{O}$.
Its ground state is dominated by the $p$-shell closed configuration,
while many low-lying excited states are difficult to be understood within the shell-model framework.
Cluster models~\cite{BRINK1970143,Suzuki:1976zz,Suzuki:1976zz2,fujiwara80,LIBERTHEINEMANN1980429,PhysRevC.29.1046,DESCOUVEMONT1987309,PhysRevC.44.306,PhysRevC.47.210,Fukatsu92,PhysRevLett.101.082502,PhysRevC.82.024312,PhysRevC.85.034315,PhysRevLett.112.152501,BIJKER2017154}
assuming $^{12}\mrm{C}+\alpha$ and $4\alpha$ structures have been suggested to describe these states.
However, it is difficult for such models to answer the questions
whether four nucleons form an $\alpha$ cluster in the sixteen-nucleon dynamics and
how the formed $\alpha$ cluster is distributed in the system,
since these models rely on an {\it a priori} assumption of the cluster structures.

Very recently, fully microscopic calculations based on the antisymmetrized molecular dynamics
(AMD)~\cite{KanadaEn'yo:2012nw,PhysRevC.89.024302,PhysRevC.96.034306} and
the chiral nuclear effective field theory~\cite{PhysRevLett.112.102501} have shown the formation of cluster structures
in the sixteen-nucleon system $^{16}\mrm{O}$, starting from nucleon degrees of freedom.
In addition, five-body calculations~\cite{PhysRevC.89.011304} using the $^{12}\mrm{C}+ppnn$ orthogonality condition model (OCM)
have indicated $\alpha$ cluster formation at the nuclear surface.
This five-body model (5BM) is more sophisticated than the conventional $^{12}\mrm{C}+\alpha$-OCM~\cite{Suzuki:1976zz,Suzuki:1976zz2}
(we refer to this model as just OCM). It describes $\alpha$ cluster dissociation in the inner region,
although in the 5BM $^{12}\mrm{C}$ is modeled by the $0p_{3/2}$ closed configuration ignoring excitations,
which are instead taken into account in OCM calculations.
As a result, the 5BM leads to a suppression of the $\alpha$-cluster formation in the inner region,
and therefore $\alpha$ distributions in the $0_1^+$ and $0_2^+$ states of $^{16}\mrm{O}$ are
qualitatively different from those obtained by the OCM; surface peaks of the $\alpha$
probabilities are shifted outward in both states.
Both models are, in principle, based on semimicroscopic calculations using phenomenological $^{12}\mrm{C}$-$\alpha$ potentials.

In recent years, the structure of the $4^+$ states in $^{16}\mrm{O}$ has been attracting much attention.
By OCM calculations, the $0^+_2$ (6.06~MeV), $2^+_1$ (6.92~MeV), and $4^+_1$ (10.36~MeV) states
have been identified as members of the $^{12}\mrm{C}(0^+_1)+\alpha$-cluster band~\cite{Suzuki:1976zz,Suzuki:1976zz2}.
Likewise, the AMD has predicted for the $4^+_1$ and $4^+_2$ (11.10~MeV) states the $^{12}\mrm{C}+\alpha$- and
tetrahedral 4$\alpha$-cluster structures, respectively, with significant mixing between them~\cite{KanadaEn'yo:2012nw,PhysRevC.96.034306}.
This finding is consistent with the observed $\alpha$-decay widths~\cite{AJZENBERGSELOVE19711,AJZENBERGSELOVE19771},
strong $\alpha$-transfer yields~\cite{BECCHETTI1980336,ELLIOTT1985208},
and weak two-nucleon transfer cross sections~\cite{PhysRevC.2.1271,LOWE1972323}.
However, latest predictions by Bijker and Iachello~\cite{PhysRevLett.112.152501,BIJKER2017154},
who have revived the algebraic approach of $4\alpha$ system~\cite{PhysRev.52.1083}, have lead to different results.
They have attributed the $0^+_1$, $3^-_1$ (6.13~MeV), and $4^+_1$ states to the tetrahedral-$4\alpha$ band
associated with the $T_d$ symmetry, and the $4^+_2$ state to a vibration mode on it.

In order to verify the presence of $\alpha$-cluster states, $\alpha$-transfer reactions are useful,
as they can produce nuclei in their ground states as well as in excited ones, within a consistent reaction condition.
Experimental studies of $(^6\mrm{Li},d)$ reactions on $^{12}$C and its inverse have been carried out since 1960s~\cite{PhysRev.128.2282,PhysRev.129.1750,PhysRev.134.B74,PhysRev.141.1007,PhysRev.148.1097,LOEBENSTEIN1967481,MEIEREWERT1968142,PhysRevC.9.2451,MAKOWSKARZESZUTKO1978187,BECCHETTI1978313,PhysRevC.18.856,OELERT19781,OELERT1979192,BECCHETTI1980336,GLOVER1981469,PhysRevC.59.2574,PhysRevLett.83.4025,PhysRevC.67.044604,BELHOUT2007178,ADHIKARI2011308,PhysRevC.89.044618}.
In most of these studies, spectroscopic factors (SFs) have been extracted to identify cluster states of $^{16}\mrm{O}$,
with an astrophysical interest as well.
However, such SFs exceeds unity due mainly to the uncertainty of reaction models.
Therefore, relative SFs with respect to that of the $2_1^+$ state have been regarded as an indication
to verify the $\alpha$-cluster structure, although they cannot help argue the spatial manifestation of the cluster.
Moreover, the $4^+_2$ state has been paid attention to because $\alpha$-transfer reactions such as
$^{12}\mrm{C}(^6\mrm{Li},d)^{16}\mrm{O}$ and $^{12}\mrm{C}(^7\mrm{Li},t)^{16}\mrm{O}$ anomalously yield $4^+_2$-cross sections larger
by almost two orders of magnitude than predictions evaluated from the $\alpha$-decay width~\cite{PhysRevC.9.2451,PhysRevC.14.491,PhysRevC.18.856,GLOVER1981469}.

With the aim of understanding cluster structures, $\alpha$-cluster probability in the surface and outer regions, in particular,
the surface-peak position of $\alpha$-cluster wave functions should be extracted
from reaction observables instead of SFs given as integrated quantities.
In Ref.~\cite{PhysRevC.93.034606}, we have reported that $\alpha$-transfer reactions are suitable to extract
the $\alpha$-cluster probability at the surface in the ground state of $^{20}\mrm{Ne}$.
In addition to $\alpha$-transfer reactions, also proton-induced $\alpha$-knockout reactions~\cite{PhysRevC.94.044604,PhysRevC.97.044612,PhysRevC.98.024614} and $^{12}\mrm{C}+\alpha$ inelastic scattering~\cite{PhysRevC.97.044608} have been applied to extract
the $\alpha$-cluster probability.

In this paper, our purpose is to determined the surface-peak position in the ground and excited states of $^{16}\mrm{O}$,
through an analysis of the $\alpha$-transfer reaction $^{12}\mrm{C}(^6\mrm{Li},d)^{16}\mrm{O}$.
To this end, a phenomenological potential model (PM) is introduced.
We test the $\alpha$-cluster wave functions of the $0^+_1$ and $0^+_2$ states
computed microscopically with the OCM~\cite{Suzuki:1976zz,Suzuki:1976zz2} and 5BM~\cite{PhysRevC.89.011304}.
Furthermore, we discuss what we can learn from the $\alpha$-transfer cross sections about structures
of the $4^+_1$ and $4^+_2$ states, and try to give an answer to their controversial assignments.
To avoid complicated reaction mechanism originating from compound-nucleus formation
and to be consistent with parameterization of a $^6\mrm{Li}$-optical potential~\cite{PhysRevC.13.648} we adopt,
the experimental data of the transfer reaction at 42.1~\cite{BECCHETTI1978313} and 48.2~MeV~\cite{BELHOUT2007178} are analyzed.

This paper is organized as follows.
Section~\ref{theory} is dedicated to sketch our theoretical framework, namely the model setting.
In Sec.~\ref{result}, we show results for the $0^+$ and $4^+$ states, and
discuss how the spatial manifestation is probed through transfer-cross sections.
A summary is given in Sec.~\ref{summary}.

\section{Theoretical framework}
\label{theory}
The $\alpha$-transfer reaction $^{12}\mrm{C}(^6\mrm{Li},d)^{16}\mrm{O}$ is described by
the finite-range distorted-wave Born approximation (DWBA) \cite{PhysRev.133.B3,SATCHLER19641}.
As an input of the DWBA calculations, the $^{12}\mrm{C}$-$\alpha$ relative wave function $\phi_l$,
with the relative orbital angular momentum $l$, is taken from Refs.~\cite{Suzuki:1976zz,Suzuki:1976zz2}
for an OCM-based wave function, while the 5BM-based one is also taken from Ref.~\cite{PhysRevC.89.011304}.
Both the OCM and 5BM provide the reduced-width amplitude (RWA) as the $^{12}\mrm{C}$-$\alpha$ wave function.
In order to make the RWA an appropriate input for reaction calculations, we should normalize its norm to unity.
This normalized RWA is suitable to discuss the spatial manifestation of the wave function and corresponding diffraction pattern
of the cross section, rather than the SF and absolute value of the cross section.
Moreover, since the asymptotic form of the RWA by the OCM was not reported in Refs.~\cite{Suzuki:1976zz,Suzuki:1976zz2},
we connect it with the Whittaker function at $r=6$~fm,
where $r$ is the relative distance between the clusters.

In addition to the microscopic wave functions, we introduce wave functions
simulated with the phenomenological PM~\cite{PhysRevC.93.034606} based on a Woods-Saxon potential,
in order to clarify how they are probed through the transfer-cross sections.
The parameters of the Woods-Saxon potential, given in Tables~\ref{tabPM0} and~\ref{tabPM4} in Sec.~\ref{result},
are chosen to make the PM-wave functions as references for the microscopic ones (see Sec.~\ref{result} for more details).
The depth of the potential is adjusted to reproduce the $\alpha$ separation energy of $^{16}\mrm{O}$.
The PM-wave functions of both the $0_1^+$ and $0_2^+$ states are obtained using the experimental binding energy~\cite{TILLEY19931},
whereas we approximate the $4^+$-resonance states as bound states having a binding energy of $0.01$~MeV.
We adopt experimental $Q$-values of each $4^+$ state in the DWBA calculation.

The distorted wave in the initial (final) channel is calculated with the optical potential
including the Coulomb interaction forming a uniformly charged sphere potential
by Ref.~\cite{PhysRevC.13.648} (Ref.~\cite{PhysRevC.74.044615}).
For simplicity we disregard the intrinsic spin of nuclei, and apply the no-recoil limit~\cite{PhysRevC.93.034606}
to the Hamiltonian associated with the distorted wave in the final channel.
The $^6\mrm{Li}$ wave function is computed by the $\alpha$-$d$ model~\cite{PhysRevC.93.034606,doi:10.1143/PTP.125.1193}
with the two-range Gaussian interaction~\cite{doi:10.1143/PTPS.89.136}, which parameters are chosen to describe
the ground, $1^+$-, $2^+$-, and $3^+$-resonance states.

Since a main source of the theoretical uncertainty originates from the $^6\mrm{Li}$-optical potential,
we have tested another potential~\cite{PhysRevC.59.2574}, which has the volume absorption and
parameters determined at an incident energy of 50~MeV.
Note that the parameters of the surface-absorption potential~\cite{PhysRevC.13.648} is set at incident energies from 4 to 63~MeV.
As a result of numerical calculations, we have confirmed that the different choice of the $^6\mrm{Li}$-optical potential
does not change essentially the angular distribution of the transfer-cross sections,
and thus our conclusion drawn from the DWBA calculations does not depend on the $^6\mrm{Li}$-optical potential.
Therefore, in Sec.~\ref{result}, we report the results obtained with the surface-absorption potential only.

In conclusion of this section, we mention the role of the excitation of the projectile $^6\mrm{Li}$ into continuum states.
As discussed in Refs.~\cite{PhysRevC.93.034606,doi:10.1143/PTP.125.1193,PhysRevC.91.014604}, in $(^6\mrm{Li},d)$ reactions,
the effect of the continuum excitation of $^6\mrm{Li}$ can be effectively taken into account
by a $^6\mrm{Li}$-optical potential that appropriately describes elastic scattering of $^6\mrm{Li}$.
The optical potential~\cite{PhysRevC.13.648} we adopt was parameterized through comparison with experimental data of
$^6\mrm{Li}$-elastic scattering on $^{12}\mrm{C}$ at several incident energies.
Therefore, we expect that using this optical potential ensures effective prescription for the continuum excitation of $^6\mrm{Li}$.

\section{Results and discussion}
\label{result}
\subsection{$0^+$ states}
\label{result1}
\begin{table}[!b]
\begin{center}
 \caption{The radius parameter $r_0$ and diffuseness parameter $a_0$ of the Woods-Saxon potential used in the PM
 to obtain the $0^+$-wave functions.}
 \begin{tabular}{c|ccc|cc}
  \hline
  \hline
             & \multicolumn{3}{c|}{$0_1^+$}  & \multicolumn{2}{c}{$0_2^+$} \\
  \cline{2-6}
             & PM-OCM & PM-5BM & PM-mid & PM-OCM & PM-5BM \\
  \hline
  $r_0$~(fm) & 1.000  & 1.625  & 1.250  & 1.375  & 1.625 \\
  $a_0$~(fm) & 0.520  & 0.845  & 0.780  & 0.715  & 0.845 \\
  \hline
  \hline
 \end{tabular}
 \label{tabPM0}
\end{center}
\end{table}
\begin{figure}[!t]
\begin{center}
\includegraphics[width=0.60\textwidth,clip]{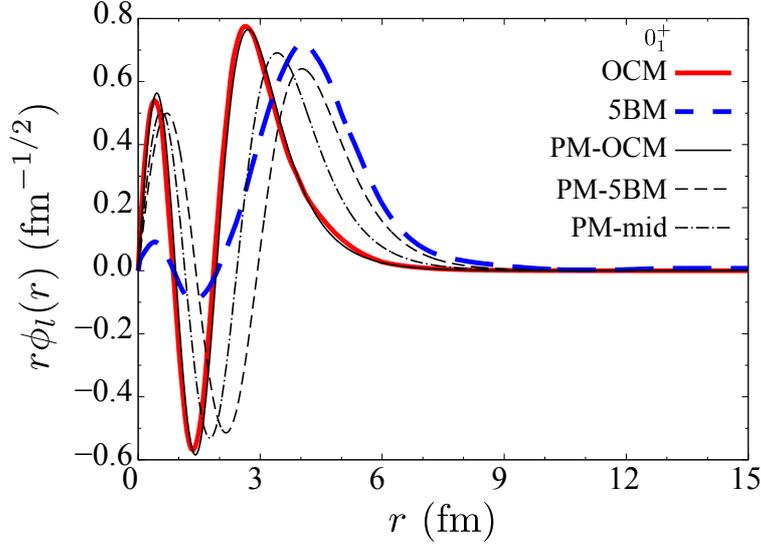}
 \caption{The $^{12}\mrm{C}$-$\alpha$ relative wave functions of the ground state of $^{16}\mrm{O}$ with $l=0$
 calculated by the OCM (thick-solid line), 5BM (thick-dashed line), PM-OCM (thin-solid line), PM-5BM (thin-dashed line),
 and PM-mid (dash-dotted line). The norm of each wave function is unity.}
\label{fig_wf01}
\end{center}
\end{figure}
\begin{figure}[!t]
\begin{center}
\includegraphics[width=0.60\textwidth,clip]{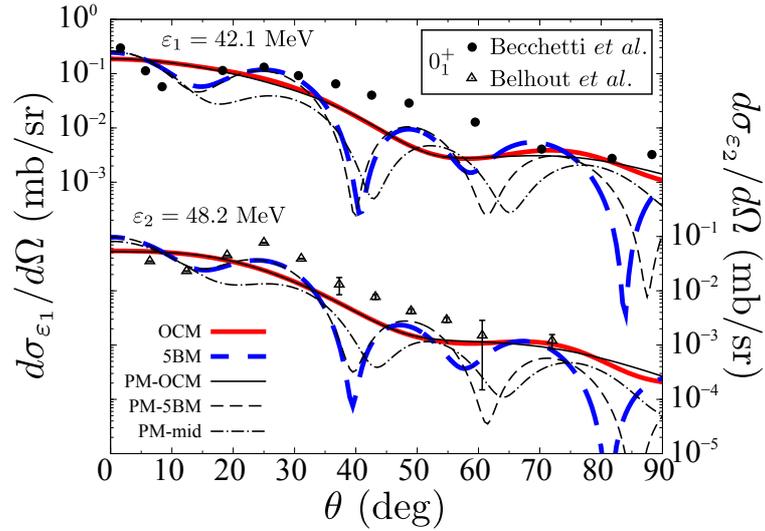}
 \caption{Comparison of the cross sections of $^{12}\mrm{C}(^6\mrm{Li},d)^{16}\mrm{O}(0_1^+)$ at
 42.1~MeV ($\varepsilon_1$) and 48.2~MeV ($\varepsilon_2$) calculated using
 $\phi_l$ of the OCM (thick-solid line), 5BM (dashed line), and PM (thin lines)
 with the experimental data~\cite{BECCHETTI1978313,BELHOUT2007178}.}
\label{fig_cs01}
\end{center}
\end{figure}
Figure~\ref{fig_wf01} represents the $^{12}\mrm{C}$-$\alpha$ relative wave function of the $0_1^+$ state of $^{16}\mrm{O}$ ($l=0$)
as a function of $r$.
The thick-solid and thick-dashed lines are taken from the Refs.~\cite{Suzuki:1976zz,Suzuki:1976zz2} for the OCM
and Ref.~\cite{PhysRevC.89.011304} for 5BM, respectively, but we normalized their norms to unity.
The wave function of the 5BM, compared to that of the OCM, has a surface peak spreading outward by $\sim 1.5$~fm
and small amplitude in the inner region, $r\lesssim 3$~fm.
In addition to the microscopic wave functions, the PM-wave functions are shown.
They are obtained by employing the parameters reported in Table~\ref{tabPM0},
where the radius and diffuseness parameters of the Woods-Saxon potential are given by $12^{1/3}r_0$ and $a_0$, respectively.
The wave function of the PM-OCM (PM-5BM) drawn with the thin-solid line (thin-dashed line) is
calculated by making its peak position at $r\sim 2.5$~fm ($r\sim 4$~fm) consistent with that of the OCM (5BM).
We also adopt the PM-mid since it is not easy to clarify how the diffraction pattern of the transfer-cross section is sensitive
to the surface-peak position of the wave function with the PM-OCM and PM-5BM only.
The PM-mid wave function, plotted by the dash-dotted line, has the surface-peak position at the middle of the OCM and 5BM ones.

In Fig.~\ref{fig_cs01}, we compare the theoretical cross sections of $^{12}\mrm{C}(^6\mrm{Li},d)^{16}\mrm{O}(0_1^+)$
as a function of the deuteron emitting angle $\theta$ in the center-of-mass frame with the experimental data
at the two incident energies, $\varepsilon_1=42.1$~MeV~\cite{BECCHETTI1978313} and $\varepsilon_2=42.8$~MeV~\cite{BELHOUT2007178}.
The thick solid, thick dashed, thin solid, thin dashed, and dash-dotted lines are the results obtained using $\phi_l$ of
the OCM, 5BM, PM-OCM, PM-5BM, and PM-mid, respectively.
The calculated results are normalized by using the normalization factor $N_0$ listed in Table~\ref{tabN00},
which are extracted from the $\chi^2$ fit to the experimental data.
When comparing theoretical and experimental data, we focus on the forward-angle region, namely,
the first and second peaks and the first dip between the two peaks of the cross section.
Then, we extract information on the $\alpha$-wave function from the position $\theta$ of the peaks and dip,
as well as from the ratio of the first peak to the second peak.

At both the incident energies, the 5BM-cross section has the peak and dip positions
consistent with those of the measured data at the forward-angle region, $\theta\lesssim 30^\circ$,
whereas the OCM gives a smooth diffraction pattern with the first dip at $\theta\sim 50^\circ$
and completely fails to explain the data.
This indicates that the shift of the surface peak of $\phi_l$ arising from the $\alpha$-cluster breaking in the 5BM
is essential to describe the ground state of $^{16}\mrm{O}$.
Our result supports that the wave function having the surface peak at $\sim 4$~fm
is eligible to describe the $0_1^+$ state, as predicted by the 5BM.

The PM results make it clear how the surface peak of the wave function is probed through the cross section at the forward angles.
Even though the shape of the PM-5BM and 5BM wave functions are significantly different from each other
in the inner region (see Fig.~\ref{fig_wf01}),
their cross sections are almost identical at the forward angles, $\theta\lesssim 50^\circ$.
Furthermore, as the surface peak is populated inward by the PM-mid from that by the PM-5BM,
the second peak of the cross section at $\theta \sim 25^\circ$ decreases and its first dip is shifted backward.
From the above results on the $0_1^+$ states, we find that the surface peak of the wave function is crucial
to describe the diffraction pattern of the cross section, and hence, not the inner region of the wave function
but its surface is sensitive to the cross section at the forward angles.
Indeed, we confirm numerically that the wave function at the inner region is absorbed by the imaginary part of the optical potentials.
The surface-peak position of the $0_1^+$-wave function can be determined by focusing
on the ratio of the first to second peaks and the position of the first dip of the cross section.
\begin{table}[!t]
\begin{center}
 \caption{The normalization factor $N_0$ for the $0^+$ states obtained through the $\chi^2$ fit of the theoretical cross sections
 to the experimental data at the two incident energies, $\varepsilon_1$ and $\varepsilon_2$.}
 \begin{tabular}{ll|ccccc}
  \hline
  \hline
                         &                 & OCM   & 5BM   & PM-OCM & PM-5BM & PM-mid \\
  \hline
  \multirow{2}*{$0_1^+$} & $\varepsilon_1$ & 1.455 & 1.494 & 1.873 & 3.040  & 6.617 \\
                         & $\varepsilon_2$ & 0.532 & 0.600 & 0.708 & 1.295  & 3.032 \\
  \hline
  \multirow{2}*{$0_2^+$} & $\varepsilon_1$ & 1.499 & 1.035 & 2.593 & 0.831 & \multirow{2}*{---} \\
                         & $\varepsilon_2$ & 0.617 & 0.297 & 1.430 & 0.238 &                    \\
  \hline
  \hline
 \end{tabular}
 \label{tabN00}
\end{center}
\end{table}

\begin{figure}[!t]
\begin{center}
\includegraphics[width=0.60\textwidth,clip]{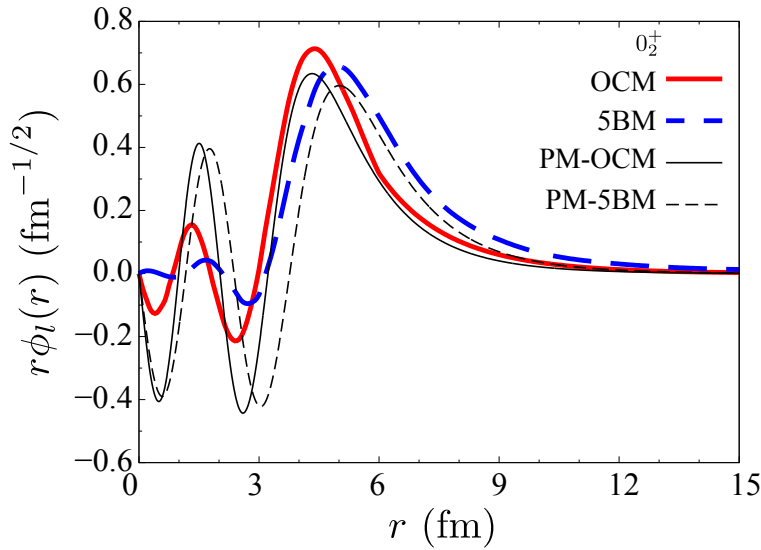}
 \caption{Same as Fig.~\ref{fig_wf01} but for the $0_2^+$ state.}
\label{fig_wf02}
\end{center}
\end{figure}

Now we show the results of the $0_2^+$ state of $^{16}\mrm{O}$.
In Fig.~\ref{fig_wf02} each line is the same as that in Fig.~\ref{fig_wf01} but for the $0_2^+$ state.
The peak position of the wave function of the 5BM at $r\sim 5$~fm,
is slightly shifted outward from that of the OCM, while in the inner region, $r \lesssim 3$~fm,
their amplitude is suppressed compared to the surface-peak amplitude.
We calculate the two PM-wave functions the surface peak of which is consistent with that of the microscopic models.
\begin{figure}[!t]
\begin{center}
\includegraphics[width=0.60\textwidth,clip]{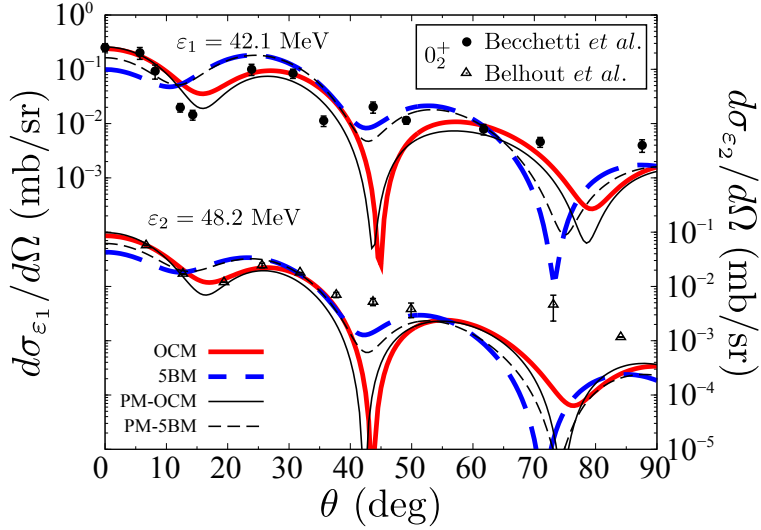}
 \caption{Same as Fig.~\ref{fig_cs01} but for the $0_2^+$ state.}
\label{fig_cs02}
\end{center}
\end{figure}

Figure~\ref{fig_cs02} represents the comparison of the calculated cross sections with the experimental data.
The legends stand for the same as those in Fig.~\ref{fig_cs01} but now for the $0_2^+$ state.
The lines are normalized to the measured data by the $\chi^2$ fitting, which results in $N_0$ given in Table~\ref{tabN00}.
At both the incident energies, the OCM adequately describes the measured data at the forward angles, $\theta\lesssim40^\circ$,
where the first peak, second peak, and first dip are explained.
In contrast, the 5BM does not account for the experimental diffraction pattern.
In particular, it is crucial that the thick-dashed line has the first peak of the cross section at the $\theta=0^\circ$
smaller than the second peak at $\theta\sim25^\circ$ for $\varepsilon_1$, and its dip is shifted forward.
It suggests that the $0_2^+$-wave function has the surface peak at $\sim 4.5$~fm predicted by the OCM,
and the 5BM does not provide appropriately the $\alpha$ probability of the $0_2^+$ state at the surface,
which is sensitively affected on the transfer-cross section.

Although, on the PM-wave functions, their amplitude in the inner region is significantly larger
than that of the microscopic-wave functions,
each PM produces the cross sections almost same as those by corresponding microscopic models,
i.e., the transfer reactions are peripheral as in the $0_1^+$ case.
The main difference between the PM-OCM and PM-5BM on the wave function can be seen at the surface-peak position,
which essentially determines the cross sections.
By comparing the cross sections of the PM-OCM and PM-5BM, we see that
the ratio of the first peak to the second peak becomes smaller and the first dip moves forward,
with manifestation of the surface peak on the wave function, as clarified for the $0_1^+$ state.
Thus we find that the inspection of the cross section at the forward angles
enables us to identify the surface-peak position of the $\alpha$-wave function of the $0_2^+$ states.

To draw a conclusion from the $0^+$ results, we recall features of the two microscopic models.
The OCM describes $^{16}\mrm{O}$ with the $^{12}\mrm{C}+\alpha$ configuration,
where the core state $^{12}\mrm{C}(0^+)$ is calculated based on the mixing of the $0p_{3/2}$-subshell-closed configuration
and the 3$\alpha$ configuration, involving the excitation of $^{12}\mrm{C}$ as well.
The 5BM addresses a dynamical process of $\alpha$ clusters by the four-nucleon correlation,
which induces the dissociation (manifestation) of $\alpha$ particles at the interior (exterior) of $^{16}\mrm{O}$,
though $^{12}\mrm{C}$ is assumed to have the $0p_{3/2}$-subshell-closed configuration only.
Our DWBA analysis for the $0_1^+$ state supports not the OCM-wave function but that of the 5BM.
This is because the $\alpha$ cluster is hard to form at the surface owing to its dissociation,
and hence its probability is shifted outward.
For the $0_2^+$ state, however, the OCM-wave function rather than that of the 5BM is reasonable to account for the transfer reaction.
For further clarification, it is desired to perform a calculation that addresses simultaneously
the $\alpha$ dissociation and core polarization of $^{12}\mrm{C}$.

As a summary of this subsection, we comment on the normalization factor $N_0$ in Table~\ref{tabN00},
where two features are found; (i) some of them exceed unity and (ii) they strongly depend on the incident energy for each $0^+$ state.
As we argued the former point (i) in Ref.~\cite{PhysRevC.93.034606},
the normalization factors extracted from our DWBA calculations are not necessarily same as physical SFs,
because the transfer reactions we analyze in this work probe only the surface region of the $^{12}\mrm{C}$-$\alpha$ wave function,
and thus we may need an artificial enhancement in order to increase the tail amplitude of the wave function.
To clarify the origin of the latter fact (ii), in future, a systematic analysis of $(^6\mrm{Li},d)$ reactions
at several incident energies is desirable.

\subsection{$4^+$ states}
\label{result2}
\begin{figure}[!b]
\begin{center}
\includegraphics[width=0.60\textwidth,clip]{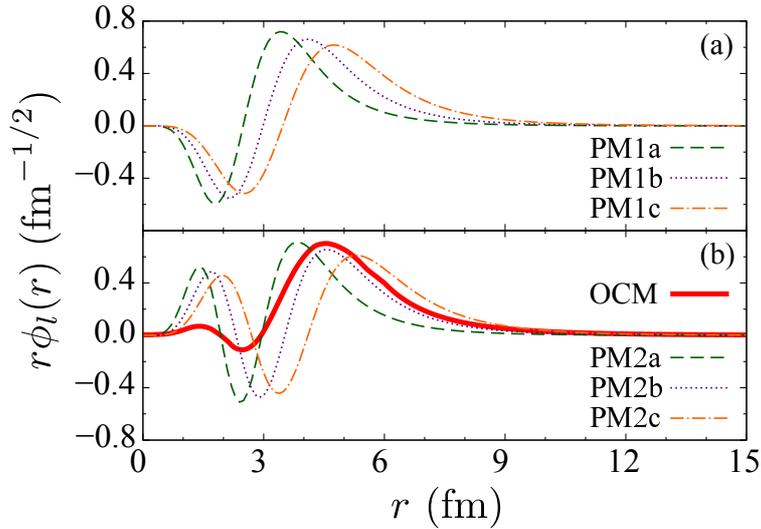}
 \caption{The $l=4$ wave functions, which norm is unity, having (a) one node and (b) two nodes
 calculated by the OCM (thick-solid line) and PM (thin lines).}
\label{fig_wf4}
\end{center}
\end{figure}
\begin{table}[!b]
\begin{center}
 \caption{Same as Table~\ref{tabPM0} but for the $4^+$ states.}
 \begin{tabular}{c|ccc}
  \hline
  \hline
             & \multicolumn{3}{c}{$4_1^+$ and $4_2^+$} \\
  \cline{2-4}
             & PM1a  & PM1b  & PM1c  \\
             & PM2a  & PM2b  & PM2c  \\
  \hline
  $r_0$~(fm) & 1.250 & 1.500 & 1.750 \\
  $a_0$~(fm) & 0.650 & 0.780 & 0.910 \\
  \hline
  \hline
 \end{tabular}
 \label{tabPM4}
\end{center}
\end{table}
In this subsection, we report the results for the analysis of the $^{12}\mrm{C}(^6\mrm{Li},d)^{16}\mrm{O}$ reaction
generating the $4_1^+$ and $4_2^+$ states of $^{16}\mrm{O}$.
First, we try to determine the surface-peak position from the angular distribution of the $\alpha$-transfer cross section,
although one sees that our calculation is not satisfactory to confirm it uniquely.
Second, we compare surface $\alpha$-probability between the $4_1^+$ and $4_2^+$ states by means of inclusive data
such as normalization factors and reduced widths.

The microscopic-wave function by the OCM is available only for the $4_1^+$ state.
As shown in Fig.~\ref{fig_wf4}(b), the OCM-wave function expressed by the solid line has a surface peak at around $4.5$~fm.
Using the parameters listed in Table~\ref{tabPM4}, we compute three sets of $\phi_l$ with $l=4$ for each $4^+$ state by the PM;
PM1a, PM1b, and PM1c (PM2a, PM2b, and PM2c) characterized by one node (two nodes).
The surface-peak position of the PM2b is tuned to coincide with that of the OCM.
Then we select the Woods-Saxon parameters in order to let the PM2a (PM2c) have the surface-peak positions inside (outside)
that of the PM2b.
The same sets of $r_0$ and $a_0$ are used for the one-node wave functions, with adjusting the depth of the potential.
The PM-wave functions are plotted by the thin lines in Fig.~\ref{fig_wf4}.
The norm of each wave function is unity.

\begin{figure}[!t]
\begin{center}
\includegraphics[width=0.60\textwidth,clip]{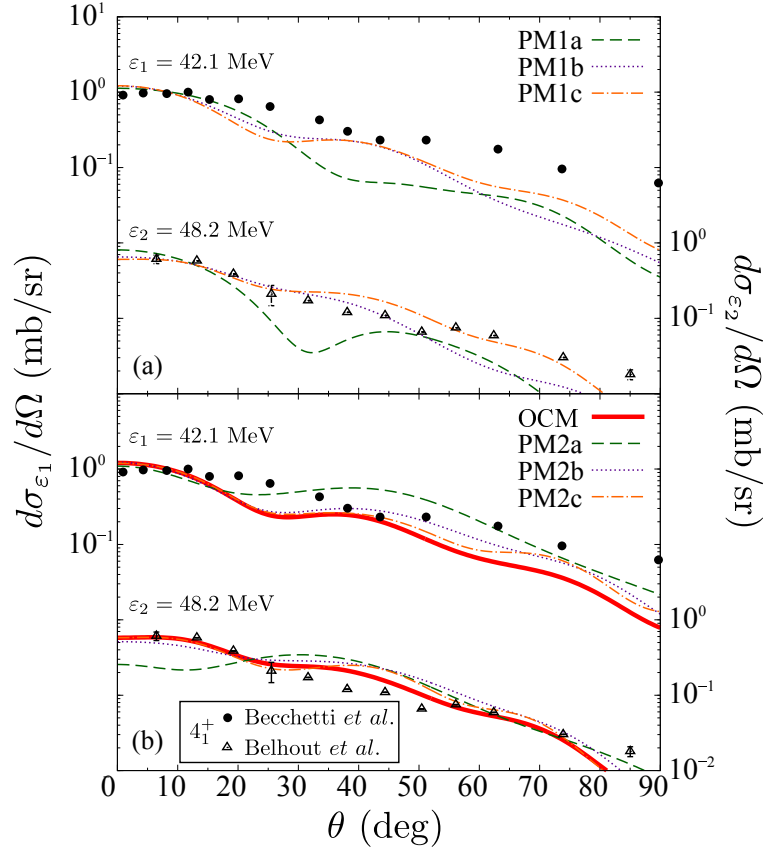}
 \caption{The angular distributed cross section of $^{12}\mrm{C}(^6\mrm{Li},d)^{16}\mrm{O}(4_1^+)$ at 42.1~MeV ($\varepsilon_1$)
 and 48.2~MeV ($\varepsilon_2$) calculated with (a) the one-node and (b) the two-node wave functions.
 In Fig.~\ref{fig_cs41}(a), the experimental data expressed by dots~\cite{BECCHETTI1978313} (triangles~\cite{BELHOUT2007178}) are same
 as those in Fig.~\ref{fig_cs41}(b).}
\label{fig_cs41}
\end{center}
\end{figure}

Figures~\ref{fig_cs41}(a) and~\ref{fig_cs41}(b) present the theoretical cross section of $^{12}\mrm{C}(^6\mrm{Li},d)^{16}\mrm{O}(4_1^+)$
obtained with the one-node and two-node wave functions, respectively.
The lines are normalized with $N_0$ found in Table~\ref{tabN04} to the experimental data
expressed by the dots and triangles for the incident energies,
$\varepsilon_1$~\cite{BECCHETTI1978313} and $\varepsilon_2$~\cite{BELHOUT2007178}, respectively.
It is difficult to extract the correspondence between the wave function and the cross section only from the result at $\varepsilon_1$,
because the diffraction pattern of the measured data at $\varepsilon_1$ is not distinct at the forward angles,
and the none of the theoretical results of both the one-node and two-node cases can explain the experimental data.
At $\varepsilon_2$, all the calculated results except the PM1a and PM2a almost identically coincide with the measured data
at forward angles, $\theta \lesssim 30^\circ$.
The PM1a at $\varepsilon_2$ produces the dip at $\sim 30^\circ$,
while the PM2a at $\varepsilon_2$ gives the first peak at $\theta = 0^\circ$ smaller than the second peak at $\theta \sim 30^\circ$.
From these results, we consider that the PM1a and PM2a are not eligible for a $4_1^+$-wave function.
Thus we confirm that the surface-peak position of $4_1^+$ is outer than $r \sim 4$~fm.
\begin{table}[!t]
\begin{center}
 \caption{The normalization factor $N_0$ for the $4^+$ states extracted by the $\chi^2$ fit of the theoretical
 cross sections to the experimental data at the two incident energies, $\varepsilon_1$ and $\varepsilon_2$.}
 \begin{tabular}{ll|ccccccc}
  \hline
  \hline
                         &                 & PM1a  & PM1b  & PM1c  & OCM                & PM2a  & PM2b  & PM2c  \\
  \hline
  \multirow{2}*{$4_1^+$} & $\varepsilon_1$ & 1.313 & 0.435 & 0.227 & 0.213              & 0.571 & 0.189 & 0.165 \\
                         & $\varepsilon_2$ & 1.243 & 0.580 & 0.337 & 0.314              & 0.395 & 0.221 & 0.198 \\
  \hline
  \multirow{2}*{$4_2^+$} & $\varepsilon_1$ & 0.374 & 0.122 & 0.065 & \multirow{2}*{---} & 0.168 & 0.055 & 0.048 \\
                         & $\varepsilon_2$ & 0.124 & 0.065 & 0.040 &                    & 0.074 & 0.031 & 0.025 \\
  \hline
  \hline
 \end{tabular}
 \label{tabN04}
\end{center}
\end{table}

\begin{figure}[!b]
\begin{center}
\includegraphics[width=0.60\textwidth,clip]{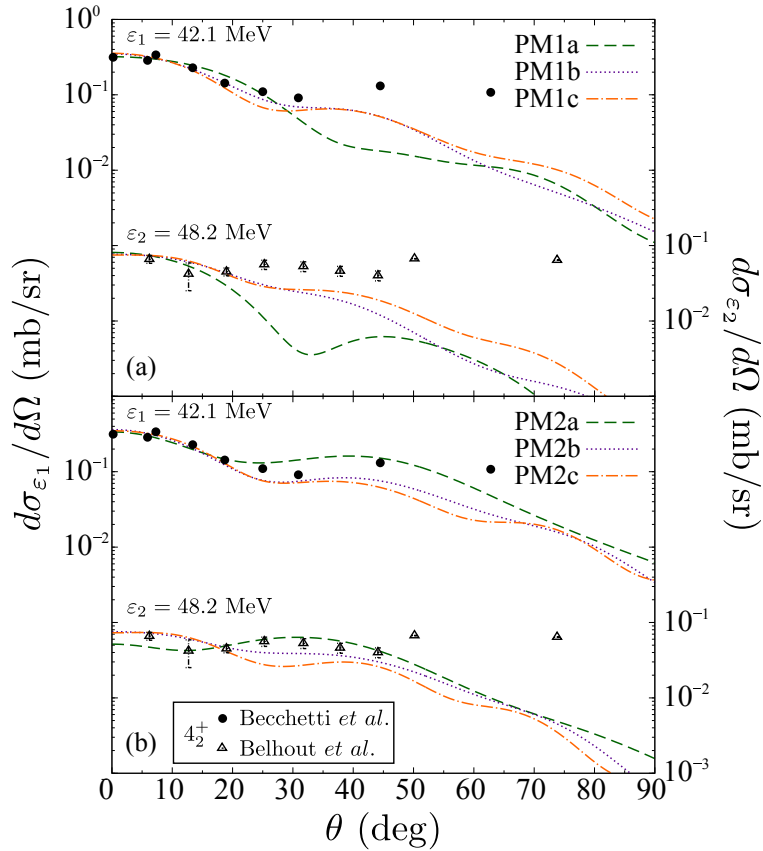}
 \caption{Same as Fig.~\ref{fig_cs41} but for the $4_2^+$ state.}
\label{fig_cs42}
\end{center}
\end{figure}

The results of the $4_2^+$ state are shown in Fig.~\ref{fig_cs42}, where the legends are the same as those in Fig.~\ref{fig_cs41}.
In Fig.~\ref{fig_cs42}(a), at both the incident energies, every line coincides with the experimental data within the region,
$\theta \lesssim 20^\circ$, even though the PM1a gives the first dip backward compared to the other PMs.
The calculations with the two-node PM in Fig.~\ref{fig_cs42}(b) reasonably explain the experimental data
of both the incident energies at $\theta \lesssim 20^\circ$, except for the PM2a, which gives the cross section
having the first peak at $\theta = 0^\circ$ smaller than the second peak at $\theta \sim 30^\circ$.
These results prevent us clarifying the spatial manifestation of the $\alpha$-cluster structure of the $4_2^+$ state
from the transfer-cross section.

It is worth noticing that, in Fig.~\ref{fig_cs42}, the measured cross sections at both the incident energies
contain the event of the $3_1^+$ (11.08~MeV) state, although it is expected to be small for the present case.
Owing to the resolution of the experiments~\cite{BECCHETTI1978313,BELHOUT2007178},
the contribution of the $3_1^+$ state being just 20~keV below the $4_2^+$ state is difficult to be separated on the cross section.
However, in Ref.~\cite{GLOVER1981469}, it was revealed that the cross section of the $3_1^+$ state at the forward angles
is only 15-20\% of the total yield for the $(3_1^+ + 4_2^+)$-doublet, within the range of the incident energy
from 20 to 34~MeV of the transfer reactions, $^{12}\mrm{C}(^6\mrm{Li},d)^{16}\mrm{O}$ and $^{12}\mrm{C}(^7\mrm{Li},t)^{16}\mrm{O}$.
Therefore we disregard the $3_1^+$ contribution in the comparison of the calculated and measured results at forward angles.

We confirm numerically that the transfer reactions populating both the $4^+$ states probe only the tail part
of the $\alpha$-wave function, due to the absorption by the imaginary part of the optical potentials.
We can also infer the peripherality because the number of nodes of the $4^+$-wave function is not verified from the cross section.
Therefore we see that the cluster component of the $4^+$ state is observed in the cross section of the transfer reaction
only through the surface region of the wave function.
In Table~\ref{tabN04}, by comparing $N_0$ between the $4_1^+$ and $4_2^+$ states within the same incident energies and the same PM,
one finds that every value of $N_0$ of the $4_1^+$ state is significantly greater than that of the $4_2^+$ state.
To understand the relation between the peripherality and the feature of $N_0$,
we compare the experimental reduced $\alpha$-width with those extracted from $N_0$ and $\phi_l$.

In Table~\ref{tabRW} we report the dimensionless reduced $\alpha$-width $\theta_l^2$ at the channel radius $a=6.0$~fm
(see \ref{SecRW} for definitions).
The experimental data listed at the rightmost column is evaluated with Eq.~\eqref{rw1}
employing the measured value of the $\alpha$-decay width $\Gamma_l$~\cite{AJZENBERGSELOVE19771},
while the other results are obtained from Eq.~\eqref{rw2} with $N_0$ and $\phi_l$
given in Table~\ref{tabN04} and Fig~\ref{fig_wf4}, respectively.
From the experimental $\Gamma_l$, the significant value of $\theta^2$ has been observed for the $4_1^+$ state,
whereas, for the $4_2^+$ state, the extremely small $\theta_l^2$ has been a long-standing puzzle since
anomalously large yields of $\alpha$-transfer cross sections of the $4_2^+$ state
were observed in spite of its small value of $\Gamma_l$~\cite{PhysRevC.9.2451,PhysRevC.14.491,PhysRevC.18.856,GLOVER1981469}.
We focus on $\theta_l^2$ calculated by the PM that reproduces the experimental diffraction pattern of the cross section
at the forward angles,
for example, the PM2b at $\varepsilon_2$ giving the values of 0.046 and 0.0064 for the $4_1^+$ and $4_2^+$ states, respectively.
The $4_2^+$ result is smaller than that of $4_1^+$ by nearly one order of magnitude,
as previously reported by the OCM~\cite{Suzuki:1976zz,Suzuki:1976zz2}.
The reduced $\alpha$-width evaluated from the cross section accounts for a characteristic of that obtained from $\Gamma_l$,
because the $4_2^+$ state has small values of $\theta_l^2$ relative to those of the $4_1^+$ state,
although it cannot explain the anomalously suppressed $\alpha$-decay width.

We draw a conclusion for the $4_2^+$ state as follows.
The $4_2^+$ state has a $^{12}\mrm{C}+\alpha$-cluster component that is not dominant,
and cannot be interpreted by a simple PM assuming the $^{12}\mrm{C}+\alpha$ configuration only,
as the AMD predicted the state mixing~\cite{KanadaEn'yo:2012nw,PhysRevC.96.034306}.
It indicates that, originating from nontrivial interference induced by the state mixing,
$\alpha$ probabilities at the surface determined by the DWBA analysis is not connected smoothly with
an asymptotic wave function responsible for the $\alpha$-decay width, within the single PM.
To simulate the $4_2^+$ structure precisely, the channel coupling between the $^{12}\mrm{C}+\alpha$ and other configurations
must be taken into account.
In other words of the direct reaction theory, it is desired to perform calculations based on the coupled-channels Born approximation,
as discussed in Refs.~\cite{GLOVER1981469,PhysRevC.9.2451,PhysRevC.18.856}.
Such advanced reaction framework may resolve the discrepancy between the $\alpha$-transfer reaction and $\alpha$-decay width.
\begin{table}[!t]
\begin{center}
 \caption{The dimensionless reduced $\alpha$-width $\theta_l^2$ at $a=6.0$~fm. The experimental value of $\theta_l^2$
 listed at the rightmost column is evaluated from the measured $\alpha$-decay width~\cite{AJZENBERGSELOVE19771}.}
 \begin{tabular}{ll|ccccccc}
  \hline
  \hline
                         &                   & PM1a   & PM1b   & PM1c  & PM2a  & PM2b   & PM2c  & Exp.                                   \\
  \hline
  \multirow{2}*{$4_1^+$} & $\!\varepsilon_1$ & 0.028  & 0.042  & 0.066 & 0.027 & 0.039  & 0.084 & \multirow{2}*{0.15\,$\pm$\,0.02}	    \\
			 & $\!\varepsilon_2$ & 0.027  & 0.056  & 0.097 & 0.019 & 0.046  & 0.10  &				            \\
  \hline
  \multirow{2}*{$4_2^+$} & $\!\varepsilon_1$ & 0.0079 & 0.012  & 0.019 & 0.0079& 0.011  & 0.024 & \multirow{2}*{0.00059\,$\pm$\,0.00020} \\
			 & $\!\varepsilon_2$ & 0.0027 & 0.0062 & 0.011 & 0.0034& 0.0064 & 0.013 &                          \\
  \hline
  \hline
 \end{tabular}
 \label{tabRW}
\end{center}
\end{table}

\section{Summary}
\label{summary}
We have investigated the spatial manifestation of the $\alpha$-cluster structure
of $^{16}\mrm{O}$ through the DWBA analysis of the $\alpha$-transfer reaction $^{12}\mrm{C}(^6\mrm{Li},d)^{16}\mrm{O}$.
It is remarkable that we have shown how much $\alpha$-cluster component spatially manifest itself
from the inspection of $\alpha$-transfer cross sections for both the ground state and excited states, without using SFs.

By testing the $^{12}\mrm{C}$-$\alpha$ relative wave functions obtained in the previous studies with the microscopic models,
the OCM~\cite{Suzuki:1976zz,Suzuki:1976zz2} and 5BM~\cite{PhysRevC.89.011304},
we have verified that the $\alpha$-cluster structure manifests itself at the radius $r\sim 4$~fm ($r\sim 4.5$~fm)
for the $0_1^+$ ($0_2^+$) state.
By introducing the phenomenological PM, we have clarified the correspondence
between the $\alpha$-wave function and the transfer-cross section.
It has been found that the $\alpha$-transfer cross section for both the $0_1^+$ and $0_2^+$ sates of $^{16}\mrm{O}$
at the forward angles probes only the surface region of the wave function,
and the surface-peak position of the wave function can be determined from the ratio of the first and second peaks
of the cross section.

We have confirmed that the $\alpha$-transfer reaction populating the $4^+$ states is peripheral.
Although we have verified that the $4_1^+$ state has the surface peak at $r \sim 4$~fm or outer,
it is difficult to uniquely determine it.
For the $4_2^+$ state, we have found that the extraction of the $\alpha$ probability from the cross sections is unfeasible.
This has evidenced the puzzle between the cross section and the $\alpha$-decay width.
Our conclusion is that the cluster component in the $4_2^+$ state is not dominant but finite,
and not only the $^{12}\mrm{C}+\alpha$ but also other configurations play a role.
To extract spatial manifestation of the $\alpha$-cluster component in the $4^+$ states, from a theoretical point of view,
calculations based on the coupled-channels Born approximation,
as well as the coupled-reaction channels and compound-nucleus processes,
are expected to be performed in future to describe $^{12}\mrm{C}(^6\mrm{Li},d)^{16}\mrm{O}(4^+)$.
From the experimental side, it is desirable to carry out measurement that makes possible to separate
the $3_1^+$ and $4_2^+$ events.

In the OCM and 5BM which are employed as the input of our DWBA calculations,
the $^{12}\mrm{C}$ core is assumed and phenomenological potentials are used.
The $^{12}\mrm{C}$-$\alpha$ potential in the OCM is given by the direct potential,
which is derived from simple Gaussian interactions with the strength phenomenologically adjusted
to the $^{16}\mrm{O}$-ground-state energy relative to the $\alpha$-decay threshold.
The direct potential is equivalent to the folding potential in a single-channel problem.
The 5BM adopts a phenomenological $^{12}\mrm{C}$-nucleon potential forming a Woods-Saxon function
and its derivative with the strength tuned to the relative energies of $^{13}\mrm{C}$ at low-lying states
from the $^{12}\mrm{C}+n$ threshold energy.
In the 5BM, the valence nucleons interact with each other via the Minnesota potential~\cite{THOMPSON197753},
which is parameterized to describe experimental data of the $s$-wave $NN$ scattering at low energies,
as well as the binding energies of $s$-shell nuclei.
Therefore, in order to obtain information on nuclear forces from reaction calculations,
other structure models starting from the nucleon degrees of freedom are necessary.
In fully microscopic calculations of $^{16}\mrm{O}$ available at the moment~\cite{PhysRevC.89.024302,PhysRevC.96.034306},
a problem remains in the reproduction of energy spectra relative to the $\alpha$-threshold energy;
thus quantitative advances in microscopic structure models are required.

\appendix
\section{Calculation of reduced \mbox{\boldmath$\alpha$}-width}
\label{SecRW}
The reduced $\alpha$-width $\gamma_l$, which represents the $\alpha$ probability at the channel radius $a$, is defined by
\begin{align}
 \gamma_l^2(a)=\frac{\Gamma_l}{2P_l(a)}.
 \label{rw1}
\end{align}
Here the Coulomb penetrability $P_l$ is given by
\begin{align}
 P_l(a)=\frac{ka}{F_l^2(ka)+G_l^2(ka)},
 \label{CoulPene}
\end{align}
with the regular and irregular Coulomb functions $F_l$ and $G_l$, respectively,
and the $^{12}\mrm{C}$-$\alpha$-relative wave number $k$, which is determined uniquely from $\alpha$-separation energies.
Using a measured value of the $\alpha$-decay width $\Gamma_l$, we can extract experimental $\gamma_l$.
Alternatively, $\gamma_l$ is also evaluated from $N_0$ and $\phi_l$ as
\begin{align}
 \gamma_l^2(a)=\frac{\hbar^2N_0}{2\mu}a\!\left|\phi_l(a)\right|^2,
 \label{rw2}
\end{align}
where $\mu$ is the reduced mass of the $^{12}\mrm{C}$-$\alpha$ system.
In Table~\ref{tabRW} of Sec.~\ref{result2}, the dimensionless reduced $\alpha$-width is evaluated by
\begin{align}
 \theta_l^2(a)=\frac{\gamma_l^2(a)}{\gamma_{\mrm{W}}^2(a)}.
 \label{DLRW}
\end{align}
with the Wigner single-particle limit,
\begin{align}
 \gamma_{\mrm{W}}^2(a)=\frac{3\hbar^2}{2\mu a^2}.
 \label{WSP}
\end{align}

\section*{Acknowledgements}
\label{acknow}
The authors thank A.~Gargano and L.~De~Angelis for constructive comments and suggestions on the manuscript.
This work was supported by Japan Society for the Promotion of Science KAKENHI with Grant Number~JP15K17662.
The numerical calculations were performed with the computer facilities at Research Center for Nuclear Physics, Osaka University.

\bibliographystyle{elsarticle-num}
\bibliography{a-transfer_16O}

\end{document}